\documentclass{IEEEtran}
\usepackage{amssymb}
\usepackage{amsfonts}
\usepackage{amsmath}
\usepackage{algorithm}
\usepackage{graphicx,subfigure,cite,multicol,multirow,diagbox,booktabs,array}

\usepackage[dvips]{color}
\usepackage{float}
\usepackage{stfloats}
\ifCLASSINFOpdf
\else
\fi
\begin{document}
\title{On Performance Loss of DOA Measurement Using Massive MIMO Receiver with Mixed-ADCs}

\author{Baihua Shi, Lingling Zhu, Wenlong Cai, Nuo Chen, Tong Shen, Pengcheng Zhu, Feng Shu, and Jiangzhou Wang,~\emph{Fellow},~\emph{IEEE}

\thanks{B. Shi, L. Zhu, N. Chen and T. Shen are with the School of Electronic and Optical Engineering, Nanjing University of Science and Technology, Nanjing 210094, China.}
\thanks{W. Cai is with the National Key Laboratory of Science and Technology on Aerospace Intelligence Control, Beijing Aerospace Automatic Control Institute, Beijing 100854, China. (e-mail: caiwenlon@buaa.edu.cn).}
\thanks{P. Zhu is with National Mobile Communications Research Laboratory, Southeast University, Nanjing, China. (emails: p.zhu@seu.edu.cn).}
\thanks{F. Shu is with the School of Information and Communication Engineering, Hainan University, Haikou 570228, China. (e-mail: shufeng0101@163.com).}
\thanks{J. Wang is with the School of Engineering and, University of Kent, Canterbury CT2 7NT, U.K. (e-mail: j.z.wang@kent.ac.uk).}
}
\maketitle

\begin{abstract}
High hardware cost and high power consumption of massive multiple-input and multiple output (MIMO) are two challenges for the future wireless communications including beyond fifth generation (B5G) and sixth generation (6G). Adopting the low-resolution analog-to-digital converter (ADC) is viewed as a promising solution. Additionally, the direction of arrival (DOA) estimation is an indispensable technology for beam alignment and tracking in massive MIMO systems. Thus, in this paper, the performance of DOA estimation with mixed-ADC structure is firstly investigated.  The Cram\'{e}r-Rao lower bound (CRLB) for this architecture is derived based on the additive quantization noise model. Eventually, a performance loss factor and the associated energy efficiency factor is defined for analysis in detail. Simulation results show that the mixed-ADC architecture can strike a good balance among performance loss, circuit cost and energy efficiency. More importantly, just a few bits (up to 4 bits) of low-resolution ADCs can achieve a satisfactory performance for DOA measurement.
\end{abstract}

\begin{IEEEkeywords}
massive MIMO, DOA estimation, mixed-ADC, CRLB
\end{IEEEkeywords}

\IEEEpeerreviewmaketitle
\vspace{-3mm}
\section{Introduction}
Direction of arrival (DOA) estimation has attracted lots of attention \cite{DOA}.
In the massive multiple-input multiple-output (MIMO) systems, DOA is a key technology for its integral role in many emerging applications, including unmanned aerial vehicle (UAV) communications,
secure and precise wireless transmission systems \cite{shuSPWT}, and millimeterwave-based massive MIMO for beyond fifth generation (B5G).

It is well-known that there are two main categories in DOA estimation methods: subspace based methods and parametric methods. Estimation of signal parameters via rotational invariance technique (ESPRIT) and multiple signal classification (MUSIC) are two most famous subspace based methods \cite{MUSIC,ESPRIT}.
Authors in \cite{shuDOA} considered DOA estimation in hybrid massive MIMO systems.
In order to tackle the phase ambiguity, a deep-learning-based method for a hybrid uniform circular array (UCA) was firstly proposed in \cite{huDLDOA}.
Then, the performance of DOA estimation for low-resolution ADC structure was investigated in \cite{Shiarxiv}.

Replacing high-resolution ADCs with low-resolution ADCs is a promising solution to reduce the high hardware cost and high power consumption.
However, it is hard to analyse the performance of nonlinear signal. \cite{SinghLowADC} showed that the additive quantization noise model (AQNM) can be applied to eliminate that distortion caused by low-resolution ADCs.
However, massive MIMO systems with pure low-resolution ADCs/DACs face some tricky challenges, such as time-frequency synchronization, channel estimation, and achievable rate \cite{Zhangjsac}. Thus, a new structure, mixed-ADC structure, was proposed in \cite{LiangMIXED} to make up the shortcomings in low-resolution structure.
The performance of this system was investigated over the Rician fading channel in \cite{Zhangjsac}.
However, the DOA estimation for a massive MIMO system with mixed-ADC is still an open challenging problem.

To the best of our knowledge, DOA estimation in mixed-ADC massive MIMO systems has not been studied. 
It is crucial to investigate the performance and the energy efficiency (EE) of DOA estimation with mixed-ADCs, since
the mixed-ADC is valuable in practical application due to its satisfactory performance and very low energy consumption.
Our main contributions are summarized as follows:
\begin{enumerate}
    \item The AQNM is adopted to establish the linear system model of the DOA estimation with mixed-ADCs in massive MIMO systems. Then, based on that model, we prove that subspace-based methods can be utilized without modification in this system.
    \item In order to assess the performance, the closed-form expression of the CRLB for mixed-ADC structure is derived. In addition, by defining the performance loss factor, the specific performance loss can be calculated through theoretical computations.
    \item Finally, we make an investigation on the EE of the mixed-ADC structure. The EE factor of DOA estimation with mixed-ADC is firstly proposed. And, by resorting to the energy consumption model, the optimal number of quantization bits for different proportions of high-resolution ADCs is given. Simulation results show that mixed-ADC architecture can achieve a better trade-off with a few bits (up to 4 bits) of  ADCs in most applications.
\end{enumerate}


\vspace{-3mm}
\section{System Model}
\begin{figure}
  \centering
  \includegraphics[width=0.43\textwidth]{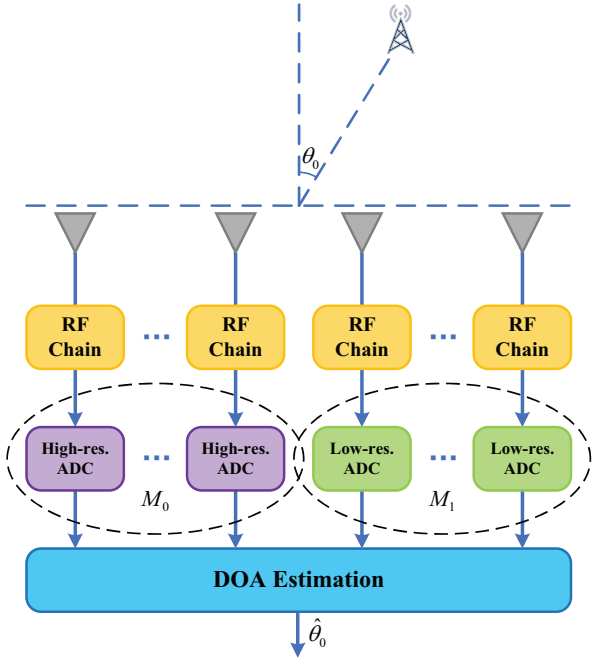}\\
  \caption{System model of the mixed-ADC massive MIMO receive array with $M_0$ high-resolution ADCs and $M_1$ low-resolution ADCs.}
  \label{sys_mod}
  \vspace{-3mm}
\end{figure}

As shown in Fig~\ref{sys_mod}, we consider a uniform linear array (ULA) equipped with mixed-ADCs. The array has $M_0$ high-resolution ADCs and $M_1$ low-resolution ADCs. The ULA has $M$ antenna elements, and we define $\kappa \triangleq M_0 / M(0\leq\kappa\leq 1)$ as the proportion of high-resolution ADCs in the mixed-ADC architecture where $M=M_0+M_1$.
The signal vector before quantization is given by
\begin{equation}\label{x}
\mathbf{x}(t)=\mathbf{a}(\theta_{0})s(t)+\mathbf{w}(t),
\end{equation}
where $\mathbf{w}(t)\sim \mathcal{CN}(0,\mathbf{I}_M)$ is the additive white Gaussian noise (AWGN), $s(t)$ is the received far-field narrow-band signal and $\mathbf{a}(\theta_{0})$ is the so-called array manifold, defined by
\begin{equation}\label{a}
\mathbf{a}(\theta_{0})=\left[e^{j 2\pi \Psi_{\theta_{0}} (1)}\ e^{j 2\pi \Psi_{\theta_{0}} (2)}\ \cdots\ e^{j 2\pi \Psi_{\theta_{0}}(M)}\right]^T,
\end{equation}
where $\Psi_{\theta_{0}}(m) = \frac{d_{m}\sin\theta_{0}}{\lambda}$
is the phase corresponding to the propagation delay of $m$th antenna element, where $\lambda$ is the wavelength of the signal.
\setcounter{equation}{7} 
\begin{figure*}[hb] 
\hrulefill
\begin{align}\label{RyM}
\mathbf{R}_{\mathbf{y}}&=\mathbb{E}[\mathbf{y}\mathbf{y}^H]
=\underbrace{\gamma
\left[
\begin{array}{cc}
  \mathbf{a}_{0}\mathbf{a}_{0}^H~~       & \mathbf{a}_{0}\mathbf{a}_{1}^H \\
  \mathbf{a}_{1}\mathbf{a}_{0}^H~~       & \mathbf{a}_{1}\mathbf{a}_{1}^H
\end{array}
\right]+\mathbf{I}_M}_{\mathbf{R}_{\mathbf{y}}'}
+\underbrace{
\left[
\begin{array}{cc}
  \mathbf{0}_{M_0}~~                                    & (\alpha-1)\gamma\mathbf{a}_{0}\mathbf{a}_{1}^H \\
  (\alpha-1)\gamma\mathbf{a}_{1}\mathbf{a}_{0}^H~~      & (\alpha^2-1)\gamma\mathbf{a}_{1}\mathbf{a}_{1}^H+(\gamma\alpha-1)(1-\alpha)\mathbf{I}_{M_1}
\end{array}
\right]}_{\mathbf{R}_{\epsilon}}
\end{align}
\setcounter{equation}{9} 
\begin{equation}\label{var}
CRLB=  \frac{3\lambda^2(\beta\gamma+1)[\beta M_0(\gamma+1)+\alpha M]+ (\beta\gamma+1)^2}{4N\pi^2\gamma\cos^2\theta_{0} d^2 [\beta M_0(\gamma+1)+\alpha M][J_0(2M_0-1)+J(2M-1)]-3(J_0+J)^2}
\end{equation}
\setcounter{equation}{10} 
\begin{align}\label{plossf}
\eta_{PL}&=\frac{\emph{CRLB}_{\kappa,\alpha}}{\emph{CRLB}_{\kappa=1,\alpha=1}}=\frac{\gamma(g+\alpha)(g\kappa+\alpha )+\frac{(\beta\gamma+1)^2}{M}}{\gamma+\frac{1}{M}}\nonumber\\
&~~~\cdot\frac{2\left(1-\frac{1}{M}\right)\left(2-\frac{1}{M}\right)-3\left(1-\frac{1}M{}\right)^2} {2(g\kappa+\alpha)\left[g\kappa\left(\kappa-\frac{1}{M}\right)\left(2\kappa-\frac{1}{M}\right) +\alpha\left(1-\frac{1}{M}\right)\left(2-\frac{1}{M}\right)\right]-3\left[g\kappa\left(\kappa-\frac{1}{M}\right)+\alpha\left(1-\frac{1}{M}\right) \right]^2}
\end{align}
\end{figure*}
\setcounter{equation}{2}

For the ease of expression and analysis, the received signal is denoted as $\mathbf{x}(t)=[\mathbf{x}_0(t),~\mathbf{x}_1(t) ]^T$,
where $\mathbf{x}_0(t)$, $\mathbf{x}_1(t)$ are the $M_0\times1$ and $M_1\times1$ vector, which denote the signals that will be quantized by high-resolution ADCs and low-resolution ADCs, respectively. Thus, the received signals experiencing the $M_0$ high-resolution ADCs can be written as $\mathbf{y}_{0}(n)=\mathbf{x}_0(t)|_{t=n}=\mathbf{a}_0(\theta_{0})s(n)+\mathbf{w}_0 (n),~n=1,2,\cdots,N,$
where $N$ is the number of snapshots. Furthermore, by leveraging on AQNM \cite{Zhangjsac},
the received signals quantized by $b$-bit ADCs can be formulated as
\begin{align}\label{yf1}
\mathbf{y}_{1}(n)=\mathbb{Q}(\mathbf{x}_{1}(t))\approx\alpha\mathbf{a}_{1}(\mathbf{\theta_{0}})s(n)+\alpha&\mathbf{w}_{1}(n)+\mathbf{w}_{q}(n),
\end{align}
where $\mathbf{w}_{q}(n)$ denotes the quantization noise, $\mathbb{Q}(\cdot)$ is the quantization function, and $\alpha=1-\beta$  is the linear quantization gain, where $\beta=\frac{\mathbb{E}\left[ \|\mathbf{x}_1-\mathbf{y}_{1}\|^{2}\right]}{\mathbb{E}\left[ \|\mathbf{x}_1\|^{2}\right]}$
denotes the distortion factor of the low-resolution ADC. The accurate value of $\beta$ is listed in Table \ref{tab1} when $b\leq5$. For longer quantization bitlength (e.g., $b>5$), the distortion factor $\beta$ can be approximated as $\beta\approx\frac{\sqrt{3}\pi}{2}\cdot 2^{-2b},~b\geq 6.$
For a fixed channel realization, the covariance matrix of $\mathbf{w}_q$ is given by
\begin{align}\label{R_wf}
\mathbf{R}_{\mathbf{w}}&=\alpha\beta \mathbf{diag}(\sigma_s^2\mathbf{a}_{1}(\theta_{0}) \mathbf{a}_{1}^H(\theta_{0})+\mathbf{I}_{M_1})\nonumber\\
&=\alpha\beta(\sigma_s^2+1)\mathbf{I}_{M_1}.
\end{align}
Thus, $\mathbf{w}_q$ can be modelled as $\mathbf{w}_q\sim \mathcal{CN}(0,\mathbf{R}_{\mathbf{w}_F})$.

\begin{table}
\footnotesize
\centering
\caption{Distortion Factors $\beta$ For Different ADCs Quantization Bits ($b\leq5$)}
\label{tab1}
\scalebox{1.15}{
\begin{tabular}{c|c|c|c|c|c}
\hline
\hline
$b$     & 1     & 2     & 3         & 4         & 5     \\
\hline
$\beta$ & 0.3634& 0.1175& 0.03454   &0.009497   &0.002499     \\
\hline
\hline
\end{tabular}}
\vspace{-3mm}
\end{table}

Thus, the overall received signal can be expressed as
\begin{align}\label{yf}
\mathbf{y}(n)
\approx
\left[
\begin{array}{c}
\mathbf{a}_{0}(\theta_{0})s(n)+\mathbf{w}_{0}(n) \\
\alpha\mathbf{a}_{1}(\theta_{0})s(n)+\alpha \mathbf{w}_{1}(n)+\mathbf{w}_{q}(n)
\end{array}
\right].
\end{align}

\section{Analysis for the Application of Root-MUSIC in Mixed-ADC Structure}
The MUSIC proposed in \cite{MUSIC} is a well-known subspace-based method for DOA estimation. In this section, we prove the fact that there is no change on MUSIC method.

When the array is equipped with pure high-resolution ADCs, the eigenvalue decomposition (EVD) of the covariance matrix $\mathbf{R}_{\mathbf{y}}'$ can be written as
\begin{align}\label{R_ss}
\mathbf{R}_{\mathbf{y}}'&=\mathbb{E}[\mathbf{y}\mathbf{y}^H]=\sigma_s^2\mathbf{a}\mathbf{a}^H+\mathbf{I}_M\nonumber\\
&=\mathbf{E}_S\mathbf{\Lambda}_S\mathbf{E}_S^H+\mathbf{E}_W\mathbf{\Lambda}_W\mathbf{E}_W^H,
\end{align}
where the columns of $\mathbf{E}_S$ and $\mathbf{E}_W$ are the eigenvectors corresponding to the useful signal and  channel noise, respectively. Those can span the signal subspace and noise subspace, which are orthogonal. Thus, the spectrum,
\begin{equation}\label{spec}
S(\theta)=\|\mathbf{E}_W^H\mathbf{a}(\theta)\|^{-2},
\end{equation}
will be infinity when $\theta=\theta_{0}$.
In practice, $\mathbf{R}_{\mathbf{y}}$ is estimated from sampled data by $\hat{\mathbf{R}}_{\mathbf{y}}=\frac{1}{N}\sum_{n=1}^{N}\mathbf{y}(n)\mathbf{y}^{H}(n).$

In mixed-ADC structure, the covariance matrix $\mathbf{R}_{\mathbf{y}}$ can be casted as (\ref{RyM}) at the bottom of next page,
\setcounter{equation}{8}
where $\gamma=\mathbb{E}[ss^H]=\sigma_{s}^2=\sigma_{s}^2 / \sigma_{n}^2$ is the input SNR of ADCs.
Observing (\ref{RyM}), as the number of quantization bit increases, $\mathbf{R}_{\epsilon}\rightarrow \mathbf{0}_M$ and $\|\mathbf{R}_{\epsilon}\|\ll\|\mathbf{R}_{\mathbf{y}}'\|$, $\mathbf{R}_{\mathbf{y}}\approx\mathbf{R}_{\mathbf{y}}'$ and $\mathbf{R}_{\epsilon}$ can be considered as the error of $\mathbf{R}_{\mathbf{y}}'$.
Thus, $\mathbf{R}_{\mathbf{y}}\approx\mathbf{R}_{\mathbf{y}}'$ and $\mathbf{R}_{\epsilon}$ can be considered as the error of $\mathbf{R}_{\mathbf{y}}'$.
Now, let $\mathbf{e}'$ and $\lambda'$ denote the eigenvalue and eigenvector of $\mathbf{R}_{\mathbf{y}}'$, respectively. Then, we have $\mathbf{R}_{\mathbf{y}}'\mathbf{e}'=\lambda'\mathbf{e}'$, which can be regarded as the linear equations $\mathbf{R}_{\mathbf{y}}'\mathbf{x}=\mathbf{b}$. And the condition number of $\mathbf{R}_{\mathbf{y}}'$ is given by
\begin{equation}\label{conditionN}
\mathrm{cond}(\mathbf{R}_{\mathbf{y}}')=\frac{\|\lambda'_{\max}(\mathbf{R}_{\mathbf{y}}')\|}{\|\lambda'_{\min}(\mathbf{R}_{\mathbf{y}}')\|}=\gamma+1.
\end{equation}
Obviously, $\mathbf{R}_{\mathbf{y}}'$ has a low condition number. And, we can conclude that $\mathbf{R}_{\mathbf{y}}'$ and $\mathbf{R}_{\mathbf{y}}$ will have the approximate eigenvalues and eigenvectors.
Thus, MUSIC and other subspace-based methods can be applied without modification, like root-MUSIC and ESPRIT \cite{ESPRIT}.

\vspace{-3mm}
\section{Performance Loss and Energy Efficiency}
In this section, to evaluate the performance of the massive MIMO for mixed-ADC architecture, we derive the CRLB for the mixed-ADC structure.
Then, the performance loss factor and EE factor are firstly defined to seek a basic trade-off between the performance and power consumption in practical applications.

\vspace{-3mm}
\subsection{CRLB and Performance Loss}
Now, let us define $d_m=(m-1)d$. And, according to Appendix, we have the CRLB for the mixed-ADC structure, (\ref{var}), as shown at the bottom of this page, where $J_0=\beta M_0(M_0-1)(\gamma+1)$ and $J=\alpha M(M-1)$.
\setcounter{equation}{11}


Furthermore, define the performance loss factor $\eta_{PL}$ by (\ref{plossf}), which is shown at the bottom of this page,
where $g=\beta(\gamma+1)$. In massive MIMO systems, when the number of antenna elements increases without bound, $M\rightarrow\infty$, the $\eta_{PL}$ will converge to
\begin{equation}\label{plossM}
\eta_{PL} \approx\frac{(g+\alpha)(g\kappa+\alpha)}{4(g\kappa+\alpha)(g\kappa^3+\alpha)-3(g\kappa^2+\alpha)^2}.
\end{equation}
Due to $0\leq\kappa\leq 1$, $\eta_{PL}$ is a linear monotonically increasing function of $\gamma$ for a fixed $\kappa$ and $b$. By contrary, $\eta_{PL}$ decreases as $\kappa$ and $b$ increase.

\vspace{-3mm}
\subsection{Energy Efficiency}

To the best of our knowledge, no one has investigated the energy efficiency for DOA estimation with mixed-ADCs. Thus, with the help of the definition in \cite{Zhangjsac}, EE of DOA estimation is defined as
\begin{equation}\label{epsilon}
\eta_{EE}=\frac{\emph{CRLB}^{-\frac{1}{2}}}{P_{total}}~~1/\mathrm{degree}/\mathrm{W},
\end{equation}
where $P_{total}$ is the total power consumption in the massive MIMO system. $\emph{CRLB}^{-1/2}$ represents the accuracy, which is the reciprocal of standard deviation lower bound and the unit of that is 1/degree. $P_{total}$ can be expressed as
\begin{align}\label{Pr}
P_{total}=P_{RF}+M_0P_{ADC,H}+M_1P_{ADC,L}
\end{align}
where $P_{RF}=P_{syc}+M(P_{LNA}+P_{mix}+P_{fil}+P_{IFA})+M_0 P_{AGC}+\rho M_1P_{AGC} $, where
$P_{syc}$, $P_{LNA}$, $P_{mix}$, $P_{fil}$, $P_{IFA}$, $P_{AGC}$, $P_{ADC_H}$ and $P_{ADC_L}$ are the power consumption values for the frequency synthesizer, LNA, mixer, the active filters, the intermediate frequency amplifier, AGC, high-resolution ADCs, low-resolution ADCs, respectively. In addition, $\rho$ is the flag function determined by the low-resolution ADC's bit, which is given by
\begin{equation}\label{c}
\rho=\left\{
\begin{array}{cc}
  0,~~ & b=1, \\
  1,~~ & b>1.
\end{array}
\right.
\end{equation}
The power consumption of the ADC can be calculated by
\begin{equation}\label{PADC}
P_{ADC}\approx\frac{3V_{dd}^2L_{min}(2B+f_{cor})}{10^{-0.1525b+4.838}},
\end{equation}
where $B$ denotes the bandwith of the signal, $V_{dd}$ is the supply voltage of converter, $L_{min}$ is the minimum channel length for the given CMOS technology, $f_{cor}$ is the corner frequency of the $1/f$ noise. (\ref{PADC}) is established for the complete class of CMOS Nyquist-rate high speed ADCs \cite{PowerEstimation}.

\vspace{-3mm}
\section{Simulation Results and Discussions}
In this section, we provide the simulation results to analyse the impact of different $\kappa$ and  $b$ on the performance loss $\eta_{PL}$.
Furthermore, the simulation of MUSIC with $\kappa=1/4$ and different $b$ is conducted, where all results are averaged over 8000 Monte Carlo realizations.
In all simulations, it is assumed that two emitters are located in $\theta_{0}^1=-45^\circ$ and $\theta_{0}^2=30^\circ$, the number of snapshots $K$ is 32 and the number of antenna elements $M$ is $128$.

\begin{figure}
  \centering
  \includegraphics[width=0.45\textwidth]{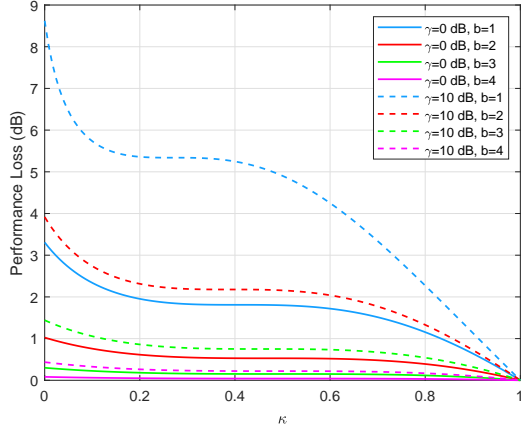}\\
  \caption{Performance comparison of the DOA estimation over $\kappa$.}
  \label{fig_pl}
  \vspace{-3mm}
\end{figure}

In Fig.~\ref{fig_pl}, performance loss over $\kappa$ is illustrated. We consider two common SNR: $\gamma=0dB$ and $\gamma=10dB$. Obviously, $\eta_{PL}$ decreases as $\kappa$ increases.
If we set 1 dB as an acceptable minimum of performance loss, 1-bit ADCs could be considered when $\kappa>0.9$. For a medium value of $\kappa$ ($0.2<\kappa<0.9$), ADCs with 2-bit or 3-bit are more suitable.
In addition, when $\kappa$ decreases to 0.2 below, 3-bit ADCs even 4-bit ADCs can be adopted to achieve a satisfactory performance. Of course, in some special cases, more or less quantization bit of ADCs should be chosen to meet the practical requirement.
\begin{figure}
  \centering
  \includegraphics[width=0.45\textwidth]{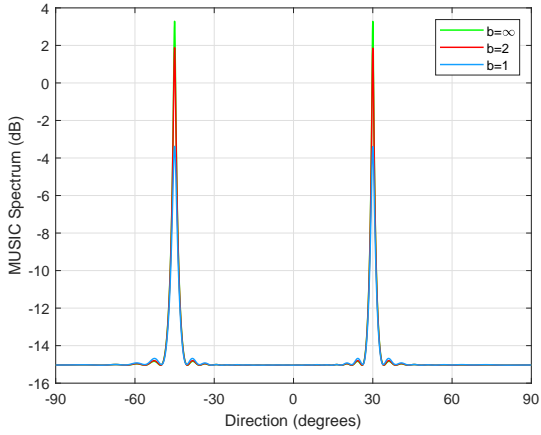}\\
  \caption{MUSIC spectrum for 32-element mixed-ADC ULA systems}
  \label{fig_MUSIC}
  \vspace{-3mm}
\end{figure}
Fig.~\ref{fig_MUSIC} plots curves of MUSIC spectrum for mixed-ADC arrays with different quantization bits. It is seen that MUSIC spectrums of arrays with 1-bit ADCs and 2-bit have lower peaks at the same directions. That proves that MUSIC method can be applied for mixed-ADC structure without modification and low-resolution ADCs have an impact on performance. An insightful conclusion is that mix-ADC structure has the similar eigen-subspace.

\begin{figure}[t]
  \centering
  \includegraphics[width=0.45\textwidth]{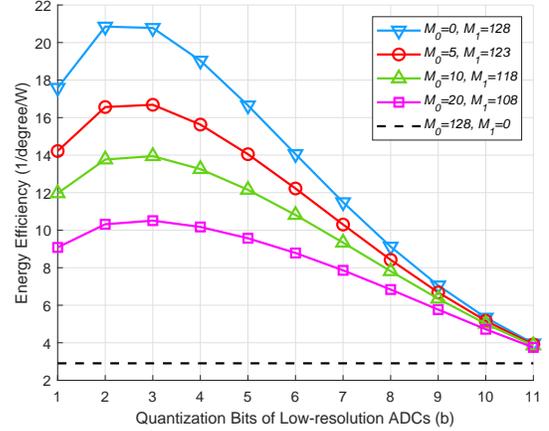}\\
  \caption{Energy Efficiency over the quantization bits of low-resolution ADCs.}
  \label{EE}
  \vspace{-3mm}
\end{figure}

The EE against the quantization bits is shown in Fig.~\ref{EE}. We adopt the classic values in massive MIMO systems: $P_{syc}=50.0~\mathrm{mW}$, $P_{LNA}=20~\mathrm{mW}$, $P_{mix}=30.3~\mathrm{mW}$, $P_{fil}=2.5~\mathrm{mW}$, $P_{IFA}=3~\mathrm{mW}$, $P_{AGC}=2~\mathrm{mW}$, $V_{dd}=3~\mathrm{V}$, $L_{min}=0.5~\mathrm{\mu m}$, $f_{cor}=1~\mathrm{MHz}$ and $B=20~\mathrm{MHz}$ as in \cite{Zhangjsac} and \cite{EnergyConstrined}. 12-bit is chosen as the quantization bit of high-resolution ADCs. It is clear that all curves reach peaks when $b=2$ or $3$. After reaching peeks, all curves decrease as the $b$ increases. And, the EE increases as $\kappa$ decreases. However, the performance of the array with pure low-resolution ADCs is too poor to be adopted.
Thus, an appropriate $\kappa$ with 1-4 bits' low-resolution ADCs is more suitable in mixed-ADC structure.
\vspace{-3mm}
\section{Conclusion}
In this paper, we built the DOA estimation for the ULA with mixed-ADC. Then, we proved that the subspace-based methods can be applied to this system without modification. Furthermore, we derived the CRLB and as the benchmark of the performance. Based on that, the performance loss and EE was investigated, which showed that 1-4 bits' ADCs are good choices in most applications. Finally, the mixed-ADC structure can achieve a satisfactory performance of DOA estimation with much less circuit cost and power consumption.

\vspace{-3mm}
\appendix[Derivation of CRLB for Full Digital Structure with Mixed-ADC]
In this section, we derive the CRLB for the massive MIMO with mixed-ADCs. In accordance with \cite{DOA}, the corresponding Fisher information matrix (FIM) $\mathbf{F}$ can be written as
\begin{equation}\label{Ff}
\mathbf{F}=\mathbf{Tr}\left\{ \mathbf{R}_{\mathbf{y}}^{-1} \frac{\partial \mathbf{R}_{\mathbf{y}}}{\partial \theta_{0}} \mathbf{R}_{\mathbf{y}}^{-1} \frac{\partial \mathbf{R}_{\mathbf{y}}}{\partial \theta_{0}} \right\}.
\end{equation}

For convenient derivation, $s(n)$, $\mathbf{y}(n)$, $\mathbf{a}(\theta_{0})$ and $\mathbf{w}(n)$ are  abbreviated as $s$, $\mathbf{y}$, $\mathbf{a}$ and $\mathbf{w}$ respectively in the following part.
Now, to simplify the derivation, we reformulate $\mathbf{y}$ as
\begin{equation}\label{yf_T}
\mathbf{y}=\mathbf{T}\mathbf{a}s+\mathbf{T}\mathbf{w}+\mathbf{q},
\end{equation}
where
\begin{equation}\label{T}
\mathbf{T}=\left[
\begin{array}{cc}
  \mathbf{I}_{M_0}              & \mathbf{0}_{M_0 \times M_1} \\
  \mathbf{0}_{M_1 \times M_0}   & \alpha\mathbf{I}_{M_1}
\end{array}
\right]
\end{equation}
and
\begin{equation}\label{qf}
\mathbf{q}=\left[
\begin{array}{c}
    \mathbf{0}_{M_0 \times 1} \\
    \mathbf{w}_{F}
\end{array}
\right].
\end{equation}
Then, the $\mathbf{R}_{\mathbf{y}}$ is given by
\begin{equation}\label{R_yf}
\mathbf{R}_{\mathbf{y}}=\mathbb{E}[\mathbf{y}\mathbf{y}^H]=\gamma\mathbf{T}\mathbf{a}\mathbf{a}^H \mathbf{T}^H+\mathbf{Q},
\end{equation}
where
\begin{equation}\label{Qf}
\mathbf{Q}=\left[
\begin{array}{cc}
  \mathbf{I}_{M_0}              & \mathbf{0}_{M_0 \times M_1} \\
  \mathbf{0}_{M_1 \times M_0}   & [\alpha^2+\alpha\beta(\sigma_s^2+1)]\mathbf{I}_{M_1}
\end{array}
\right].
\end{equation}
Thus,
\begin{equation}\label{dRf}
\frac{\partial\mathbf{R}_{\mathbf{y}}}{\partial\theta_{0}}=\gamma \mathbf{T}(\dot{\mathbf{a}}\mathbf{a}^H+\mathbf{a}\dot{\mathbf{a}}^H)\mathbf{T}^H,
\end{equation}
where $\dot{\mathbf{a}}$ is the differential of $\mathbf{a}$ to $\theta_{0}$, which is derived as
\begin{equation}\label{da}
\dot{\mathbf{a}}=\frac{d}{d\theta_{0}}\mathbf{a}(\theta_{0})=j\frac{2\pi}{\lambda}\cos\theta_{0}\mathbf{D}\mathbf{a},
\end{equation}
where
\begin{equation}\label{D}
\mathbf{D}=\left[
\begin{array}{cccc}
d_1     & 0             & \cdots & 0 \\
0       & d_2           & \cdots & 0 \\
\vdots  & \vdots        & \ddots & \vdots \\
0       & 0             & \cdots & d_M
\end{array}
\right].
\end{equation}
Therefor, with the help of some properties of the trace in \cite{matrixZhang}, FIM can be expressed as
\begin{align}\label{Ffe}
\mathbf{F}&=\gamma^2\mathbf{Tr}\left\{\mathbf{R}_{\mathbf{y}}^{-1} \mathbf{T}(\dot{\mathbf{a}}\mathbf{a}^H+\mathbf{a}\dot{\mathbf{a}}^H)\mathbf{T}^H\mathbf{R}_{\mathbf{y}}^{-1} \mathbf{T}(\dot{\mathbf{a}}\mathbf{a}^H+\mathbf{a}\dot{\mathbf{a}}^H)\right.\nonumber\\
&\left.~~\times\mathbf{T}^H\right\}=\gamma^2(F_a+2F_b+F_c),
\end{align}
where
\begin{equation}\label{Faa}
F_a=(\mathbf{a}^H\mathbf{T}^H\mathbf{R}_{\mathbf{y}}^{-1}\mathbf{T}\dot{\mathbf{a}})^2,
\end{equation}
\begin{equation}\label{Fbb}
F_b=(\mathbf{a}^H\mathbf{T}^H\mathbf{R}_{\mathbf{y}}^{-1}\mathbf{T}\mathbf{a})(\dot{\mathbf{a}}^H\mathbf{T}^H\mathbf{R}_{\mathbf{y}}^{-1}\mathbf{T}\dot{\mathbf{a}}),
\end{equation}
\begin{equation}\label{Fcc}
F_c=(\dot{\mathbf{a}}^H\mathbf{T}^H\mathbf{R}_{\mathbf{y}}^{-1}\mathbf{T}\mathbf{a})^2.
\end{equation}

In the following, resorting to the well-known Sherman-Morrison-Woodbury formula in \cite{matrixZhang},
we can get
\begin{equation}\label{R_inv}
\mathbf{R}_{\mathbf{y}}^{-1}=\mathbf{Q}^{-1}-\frac{\mathbf{Q}^{-1}\mathbf{T}\mathbf{a}\mathbf{a}^H\mathbf{T}^H \mathbf{Q}^{-1}}{\gamma^{-1}+\mathbf{a}^H\mathbf{T}^H\mathbf{Q}^{-1}\mathbf{T}\mathbf{a}}.
\end{equation}
Let us define
\begin{equation}\label{xi}
\xi=\mathbf{a}^H\mathbf{T}^H\mathbf{Q}^{-1}\mathbf{T}\mathbf{a}=M_0+\frac{\alpha}{\beta\sigma_s^2+1}M_1.
\end{equation}
Then, we can get
\begin{align}\label{Fasub}
\mathbf{a}^H\mathbf{T}^H&\mathbf{R}_{\mathbf{y}}^{-1}\mathbf{T}\dot{\mathbf{a}}\nonumber\\
&=\mathbf{a}^H\mathbf{T}^H \left( \mathbf{Q}^{-1}-\frac{\mathbf{Q}^{-1}\mathbf{T}\mathbf{a}\mathbf{a}^H\mathbf{T}^H \mathbf{Q}^{-1}}{\gamma^{-1}+\xi} \right)\mathbf{T}\dot{\mathbf{a}} \nonumber\\
&=j\frac{2\pi}{\lambda(\gamma \xi+1)}\cos\theta_{0}\mu,
\end{align}
where
\begin{equation}\label{mu}
\mu=\sum\limits_{m=1}^{M_0}d_{m}+\frac{\alpha}{\beta\sigma_s^2+1}\sum\limits_{m=M_0+1}^{M}d_{m}.
\end{equation}
Accordingly, substituting (\ref{Fasub}) into (\ref{Faa}), $F_a$ is given by
\begin{align}\label{Fafinal}
F_a &=(\mathbf{a}^H\mathbf{T}^H\mathbf{R}_{\mathbf{y}}^{-1}\mathbf{T}\dot{\mathbf{a}})^2 \nonumber\\
 &=- \frac{4 \pi^2}{\lambda^2 (\gamma \xi+1)^2} \cos^2\theta_{0} \mu^2.
\end{align}
Similar to the derivation of $F_a$, $F_c$ can be derived as follows,
\begin{align}\label{Fc}
F_c &= (\dot{\mathbf{a}}^H\mathbf{T}^H\mathbf{R}_{\mathbf{y}}^{-1}\mathbf{T}\mathbf{a})^2 \nonumber\\
&=\left[-j\frac{2\pi}{\lambda(\gamma \xi+1)} \cos\theta_{0}\mu \right]^2=F_a \nonumber\\
&= - \frac{4 \pi^2}{\lambda^2 (\gamma \xi+1)^2} \cos^2\theta_{0} \mu^2.
\end{align}

Now, let us focus on the derivation of $F_b$. The first part of (\ref{Fbb}) can be given by
\begin{align}\label{Fbsub1}
\mathbf{a}^H\mathbf{T}^H&\mathbf{R}_{\mathbf{y}}^{-1}\mathbf{T}\mathbf{a} \nonumber\\
&= \mathbf{a}^H\mathbf{T}^H \left( \mathbf{Q}^{-1}-\frac{\mathbf{Q}^{-1}\mathbf{T}\mathbf{a}\mathbf{a}^H\mathbf{T}^H \mathbf{Q}^{-1}}{\gamma^{-1}+\xi} \right)\mathbf{T}\mathbf{a} \nonumber\\
&= \frac{\xi}{\gamma \xi+1}.
\end{align}
And, the second part is expressed as
\begin{align}\label{Fbsub2}
\dot{\mathbf{a}}^H\mathbf{T}^H& \mathbf{R}_{\mathbf{y}}^{-1}\mathbf{T}\dot{\mathbf{a}}\nonumber\\
&= \dot{\mathbf{a}}^H\mathbf{T}^H \left( \mathbf{Q}^{-1}-\frac{\mathbf{Q}^{-1}\mathbf{T}\mathbf{a}\mathbf{a}^H\mathbf{T}^H \mathbf{Q}^{-1}}{\gamma^{-1}+\xi}\right)\mathbf{T}\dot{\mathbf{a}}\nonumber\\
&=\frac{4\pi^2}{\lambda^2}\cos^2\theta_{0}\left(\nu-\frac{\gamma\mu^2}{\gamma\xi+1}\right),
\end{align}
where
\begin{equation}\label{nu}
\nu = \sum\limits_{m=1}^{M_0}d_{m}^2+\frac{\alpha}{\beta\sigma_s^2+1}\sum\limits_{m=M_0+1}^{M}d_{m}^2.
\end{equation}
Combining (\ref{Fbsub1}) and (\ref{Fbsub2}), $F_b$ in (\ref{Fbb}) is represented by
\begin{align}\label{Fbfinal}
F_b &=(\mathbf{a}^H\mathbf{T}^H\mathbf{R}_{\mathbf{y}}^{-1}\mathbf{T}\mathbf{a})(\dot{\mathbf{a}}^H\mathbf{T}^H\mathbf{R}_{\mathbf{y}}^{-1}\mathbf{T}\dot{\mathbf{a}}) \nonumber\\
&= \frac{4\pi^2\xi}{\lambda^2(\gamma\xi+1)}\cos^2\theta_{0} \left(\nu-\frac{\gamma\mu^2}{\gamma\xi+1}\right).
\end{align}
Finally, substitute \ref{Fafinal}), (\ref{Fbfinal}) and (\ref{Fc}) into (\ref{Ffe}), which yields
\begin{align}\label{Ffinal}
\mathbf{F}&=\gamma^2(F_a+2F_b+F_c) \nonumber\\
&=\frac{8\pi^2\gamma^2}{\lambda^2(\gamma\xi+1)}\cos^2\theta_{0} (\xi\nu-\mu^2).
\end{align}
Therefore, the CRLB is given by
\begin{align}\label{CRLBfinal}
\emph{CRLB}=\frac{1}{N}\mathbf{F}
^{-1}=\frac{\lambda^2(\gamma\xi+1)}{8N\pi^2\gamma^2\cos^2\theta_{0}(\xi\nu-\mu^2)}.
\end{align}
The derivation of CRLB for the massive MIMO system with mixed-ADCs is completed.
$\hfill\blacksquare$
\vspace{-2mm}

\ifCLASSOPTIONcaptionsoff
  \newpage
\fi

\bibliographystyle{IEEEtran}
\bibliography{Mixed_ADC_cite}

\end{document}